\documentclass[10pt,aip,reprint]{revtex4-1}
\usepackage[pdftex]{graphicx,graphics}
\usepackage{amssymb,amsmath}

\begin{document}

\normalsize

\title{Classification of head-on collisions of ion-acoustic solitary waves in a plasma with cold ions and Boltzmann electrons}

\author{Yu. V. Medvedev}

\affiliation{Joint Institute for High Temperatures, Russian Academy of Sciences, Moscow, 125412 Russia}

\begin{abstract}

Head-on collisions of ion-acoustic solitary waves in a collisionless plasma consisting of cold ions and Boltzmann electrons are studied using the particle-in-cell  simulation.
It is shown that the collision of solitary waves can occur under different scenarios.
Solitary waves preserve or do not preserve their amplitudes and shapes after a collision, depending on their initial amplitudes.
The range of initial amplitudes, at which a solitary wave preserves its identity after collisions, is established.
The use of a diagram of initial amplitudes of colliding solitary waves to consider possible collision scenarios is discussed.
The characteristic regions in the diagram of the initial amplitudes corresponding to different collision scenarios are determined, and a classification of head-on collisions of ion-acoustic solitary waves in a plasma is proposed.
\end{abstract}

\maketitle

\section{Introduction}

Solitary waves in a collisionless plasma have been studied for a long time  \cite {Vedenov, Sagdeev}.
Particular attention has been given to the study of the Korteweg-de Vries (KdV) equation, whose solutions can describe solitary waves. 
In the analysis of numerical solutions of the KdV equation, it was established that solitary waves of small amplitudes interact without losing their  identities and these solitary waves were named ''solitons'' \cite {Zabusky}. 
In general, solitary waves can either preserve or not preserve their identities after a mutual head-on collision.
It can be said that solitons are a subclass of solitary waves.

  In this article, we consider ion-acoustic solitary waves and ion-acoustic solitons propagating in a collisionless plasma.
For brevity, we will refer to them as solitary waves or solitons,  omitting the word ''ion-acoustic''.
This will not lead to a misunderstanding, since other types of waves are not considered here.

Already at the dawn of research, a number of theoretical, numerical, and experimental studies of solitons and their interactions were carried out.
 Both head-on and overtaking collisions of two solitons in an electron-ion plasma were studied experimentally \cite{Ikezi}. 
 The overtaking collision of solitary  waves in a plasma with hot ions and Boltzmann electrons was simulated numerically by Sakanaka  \cite{Sakanaka}. 
  The collision of  solitons in a  plasma with negative ions was also investigated experimentally \cite{Nakamur84,Cooney,Lonngren,Nakamura99}.  
   A review of the early experiments with  solitons in plasmas was presented by Lonngren \cite{LonngrenRew}.
 
Recently, much attention has been paid to the study of the collision of solitary waves in a multicomponent plasma, in which, in addition to positive ions and electrons, negative ions, positrons or charged dust particles may be present  \cite {Verheest, Chatterjee10, El-Tantawy, Ruan, Roy, Roy14,Khaled,Chatterjee11,Ghosh11,Ghorui,Parveen}.
On the other hand, the collision of solitary waves in a two-component electron-ion plasma is also of considerable interest  \cite {Ghosh12,El-Tantawy15,Qi,Sharma,PPCF,Jenab}.
For example, the effect of a non-Maxwellian electron distribution in an  electron-ion plasma on the head-on collision of small-amplitude solitary waves is investigated with the help of the extended Poincar\'e-Lighthill-Kuo (PLK) method  \cite {Ghosh12, El-Tantawy15}.
 The collision of large amplitude  solitary waves was simulated by the particle-in-cell (PIC) method  \cite{Qi,Sharma}. 
 The analytical results obtained  by the PLK method were compared with the results of PIC  simulations of the collision of  solitary waves  \cite{Qi}. 
 It was shown that the results can differ appreciably in the case of large amplitudes. 
 A train of  solitons reflected from a wall and passing through one another was investigated using PIC simulations \cite{PPCF}. 
 The overtaking collision of  solitons in the presence of trapping effects of electrons was also studied \cite{Jenab}.

In the works mentioned, the main features of the collisions of solitary waves propagating either in the same or in opposite directions were considered, and changes in the  trajectories of the solitary waves  after the collisions were found.
In the cases considered, the solitary waves did not lose their identities.
They  restored their shapes and amplitudes some time after the collision.
In these studies, the colliding solitary waves had not very large amplitudes.
Note that it is the amplitude that is the only parameter that fully determines the properties of a solitary wave in a plasma with cold ions and Boltzmann electrons.

In the case of large amplitudes of colliding solitary waves, the question of preserving their amplitudes and shapes after a collision requires special study.
Recently, it was found that in the head-on collision of solitary waves of large amplitudes in a plasma consisting of cold ions and Boltzmann electrons, bunches of accelerated ions can form \cite{PhysCom,PhysPlaRep}.
These ions have high velocities.  They overtake that solitary wave, which propagates in the direction of their movement, and then move independently of the solitary wave.
The acceleration of ions occurs due to the energy of the colliding solitary waves. The loss of energy by a solitary wave leads to a decrease in its amplitude and a change in its shape.
Thus, solitary waves in a plasma can either preserve or not preserve their identities after a mutual head-on collision, depending on their amplitudes. 
In other words, the collisions of solitary waves can occur according to one or another scenario.

The main goals of this work are to determine the amplitude range, in which solitons exist in a plasma with cold ions and Boltzmann electrons, and to classify all possible head-on collisions of solitary waves in such a plasma.

\section{FORMULATION OF THE PROBLEM}

Each collision of solitary waves is determined by two values of their initial amplitudes.
To classify collisions of solitary waves, it is necessary to analyze the results of collisions for a sufficiently large number of cases that differ from each other by the choice of two values of the initial amplitudes.
In each case, we must solve the problem of the collision of solitary waves with given values of their initial amplitudes.
Such a problem is formulated as follows.
The head-on collision of two solitary waves  is considered in the one-dimensional kinetic approximation
 on the assumption that the plasma is collisionless. We  assume that the electrons are in equilibrium with the electrostatic field, and their density is determined by the Boltzmann law.
 The plasma motion is described by the Vlasov system of equations. In our case the normalized system of equations is 
  \begin{equation}
   \begin{aligned}
     &\frac{\partial f_i}{\partial t}+v\frac{\partial f_i}{\partial x}-\frac{\partial \varphi}{\partial x}\frac{\partial f_i}{\partial v}=0,\\
     &\frac{\partial^2\varphi}{\partial  x^2}= -(Z_i n_i - n_{e}),\\
    &n_{i}=\int\limits_{-\infty}^{\infty}\!f_{i}(x,v,t)\,dv,\qquad
n_{e}=\exp \varphi,
           \end{aligned}\label{e01}
 \end{equation}
where $f_{i}$, $n_{i}$ and $Z_{i}$ are the ion distribution function,  density and  electrical charge number, respectively; $n_{e}$ is the electron density; $\varphi$ is the  electrostatic potential; and $x,\;t$ and $v$ are the coordinate, time and velocity, respectively. 
The quantities
\begin{equation}
   \begin{aligned}
   & (T_{e0}/4\pi e^2n_{e0})^{1/2},\quad  (m_{i}/4\pi Z_{i}e^2n_{e0})^{1/2},\\
   &(Z_{i}T_{e0}/m_{i})^{1/2}, \quad n_{e0}, \quad T_{e0}/e \nonumber
\end{aligned}
\end{equation}
 are used as units of length, time, velocity, density, and potential  respectively. The ion distribution function is normalized to $n_{e0}(m_{i}/Z_{i}T_{e0})^{1/2}$. Here $e$ is the absolute value of the electron charge; $m_{i}$ is the ion mass;  $n_{e0}$ and  $T_{e0}$ are the unperturbed electron density and  temperature (in energy units), respectively. 

The system of equations (\ref {e01}) should be supplemented by the initial and boundary conditions characterizing the problem under consideration.
The initial conditions are as follows.
At the initial time $t=0$ in  the region $[-L, L]$ there  is  a collisionless electron-ion plasma with singly charged cold ions and electrons obeying the Boltzmann distribution. The plasma is spatially uniform everywhere except for two small regions where two solitary  waves are located. We will refer to them as solitary wave 1 and solitary wave 2.  At $t=0$,  solitary wave 1 has the amplitude of  potential  $\varphi_{m1}^0$, its propagation velocity is $D_1>0$ and its maximum potential is located at the point $x = a_1 <0$.
  Those quantities of solitary wave 2 are denoted by  $\varphi_{m2}^0$, $D_2<0$ and $x=a_2>0$, respectively. 
  
At the both boundaries of the plasma region $x = \pm L$, the specular reflection conditions for particles are set.   The Poisson equation is solved with zero electric field at the boundaries. The electric potential $\varphi (x, t)$ is measured from the zero value in the unperturbed region. 
In this formulation of the problem, two solitary waves propagate towards each other and at some time $t>0$ collide with each other and then diverge.
The task is to study the collision of solitary waves and its consequences.

One of the most efficient methods for solving the system of equations (\ref {e01}) is numerical simulation by the PIC method. We use this method to simulate the propagation and collision of two solitary  waves.
In our PIC simulations, the electrons obey the Boltzmann distribution while the ion component is  simulated by particles.
 The initial distributions of the quantities in the solitary waves are calculated using the technique described early \cite {Medvedev, Book}. 
These distributions are implemented in the numerical code by specifying the  coordinates and velocities of the  particles.
Solitary waves of any amplitudes formed in this way propagate without any changes until they collide.

In simulations, we distinguish two species of particles by their initial positions.
There are $l-$ions and $r-$ions.
At the initial time $t=0$, $l-$ions are located in the region $-L< x< 0$ and $r-$ions are located in the region $0< x< L$. 
Numerical simulation is performed for $L=200$. The initial positions of solitary wave 1 $x=a_1$ and solitary wave 2 $x=a_2$ are chosen  in such a way that the both solitary waves freely propagating reach the point $x=0$ simultaneously at $t=100$.
In this case, the region of the unperturbed plasma between the solitary waves is rather large, and we can verify that each solitary wave preserves its amplitude, shape and velocity of propagation until the collision.
  The number of cells is 16000, the spatial step is $\Delta x=0.025$. We set such a large number of cells for a good resolution of the waveform. The number of particles per cell is $2.5\times 10^3$.
 To solve the nonlinear Poisson equation, we use the quasilinearization method, in which the 
solution of the nonlinear equation reduces to solving a sequence of linear equations.
At each time step, the initial approximation for the potential is assumed to be the same as the solution obtained in the previous time step.
In this case, only a few iterations are required to obtain a solution of the nonlinear equation with good accuracy.

\section{Numerical results and discussion}

As already mentioned, in a plasma with cold ions and Boltzmann electrons, a head-on collision of two solitary waves, at least one of which has a sufficiently large initial amplitude, leads to a loss of the identity of one or both of the solitary waves after the collision
\cite{PhysCom,PhysPlaRep}.
The amplitude of the solitary wave decreases, and its shape changes.
  The reason for the decrease in the amplitudes of the colliding solitary waves is that the solitary waves slow down during the collision and the ion velocity profile breaks. Because of this, some of the ions accelerated in the field of the solitary wave  do not slow down in the course of time,  but, on the contrary, overtake the peak of the solitary wave and again enter the accelerating electric field.
 As a result, a bunch of fast ions with velocities ranging from 1.9 to 3.6 sound speeds is formed. The ions are accelerated at the expense of the energy of the solitary wave, and the amplitude of the  solitary wave decreases.
 
    The result of the collision of solitary waves depends on both initial amplitudes of the colliding solitary waves. 
Let us consider the characteristic scenarios of the interaction of two solitary waves in a wide range of their amplitudes.    
It is well known that solitary waves of small amplitudes can be described by the solution of the KdV equation, and such solitary waves preserve their amplitudes and shapes in mutual collisions.
Because of this, these solitary waves can be called  KdV solitons.
The use of the solution of the KdV equation to describe a solitary wave in a plasma with cold ions and Boltzmann electrons leads to an error if the potential amplitude exceeds 0.17 \cite {Medvedev, Book}.
 The error increases with increasing amplitude.        
On the other hand,  as  shown in a number of the papers cited above, solitary waves with initial amplitudes greater than 0.17 preserve their amplitudes and shapes after collisions and can also be classified as solitons.
And only when at least one of the colliding solitary waves has a sufficiently large amplitude, the collision of solitary waves leads to the loss of the identity of one or both of the solitary waves \cite {PhysPlaRep}.

Here we do not discuss the process of collision of solitary waves and  the formation of accelerated ions, because  these phenomena have already been described in detail in  Refs.\;28, 29. 
 We intend to determine the range of initial amplitudes of solitary waves, in which the identities of solitary waves are preserved after the collision, and the range of initial amplitudes, in which these identities are not preserved. 
Besides  the mentioned case of collision of KdV solitons, we can distinguish four characteristic cases.
Each of these cases is illustrated in one of the four figures below.
The figures show the potential distribution and phase planes of the ions before the collision at $t=0$ and after the collision at $t=150$.
By the time $ t = 150 $ each solitary wave reaches a stationary state and propagates without changing its amplitude and shape.
For convenience, in all figures, solitary wave 1 is indicated by arrow $\rightarrow$ and solitary wave 2 is indicated by arrow  $\Leftarrow$.
We also shifted the velocity axis of the $l-$ions relative to the velocity axis of the $r-$ions.

It is clearly seen in the figures that after the collision, the solitary waves propagate through the plasma, which remains unperturbed. 
 This means that the propagation of solitary waves before the collision did not disturb the state of the plasma.
Small oscillations can be seen behind the solitary waves  after the collision. 
 
\begin{figure}
\includegraphics[viewport= 209 528 426 784, width=217pt]{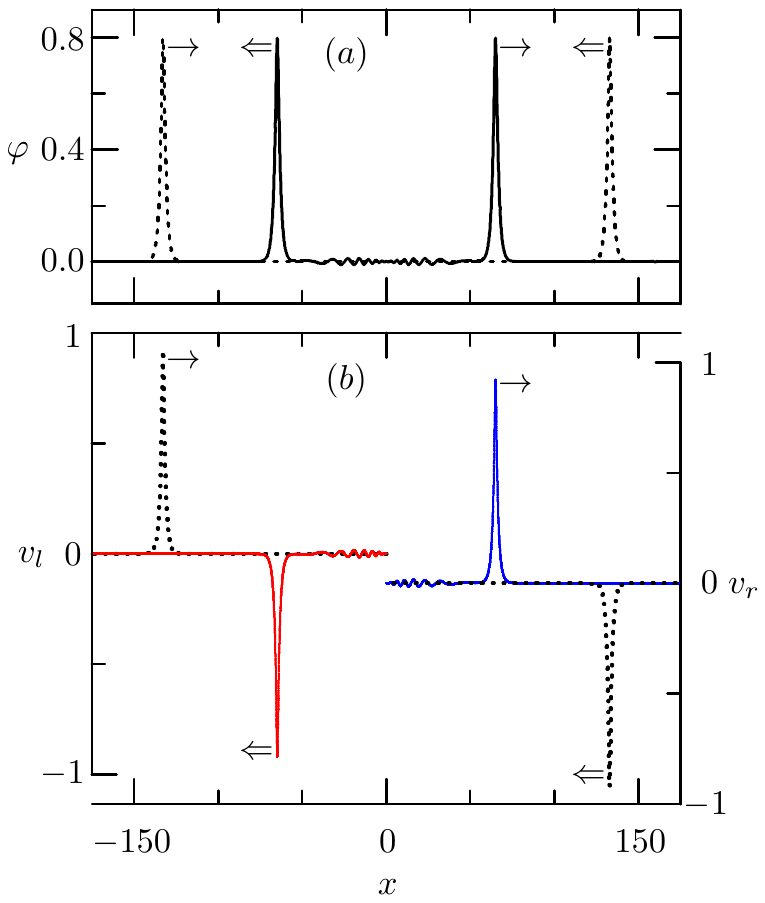}
\caption{The case of $\varphi_ {m1} ^ 0 = \varphi_ {m2} ^ 0 = 0.8$.
 $(a)$ The distribution of the potential $ \varphi (x) $ at $ t = 0 $ (dotted curves) and $ t = 150 $ (solid curves).  $(b)$
 The phase planes of $l-$ions $(x, v_l)$ (the left curves and the left axis) and the phase planes of $r-$ions $(x, v_r)$ (the right curves and the right axis) at $t = 0$ (dashed curves) and at $t = 150$ (dots, red on-line for $l-$ions and blue on-line for $r-$ions).    
}
\label{Fig1}
\end{figure}

\begin{figure}\centering
\includegraphics[viewport= 209 528 424 784, width=215pt]{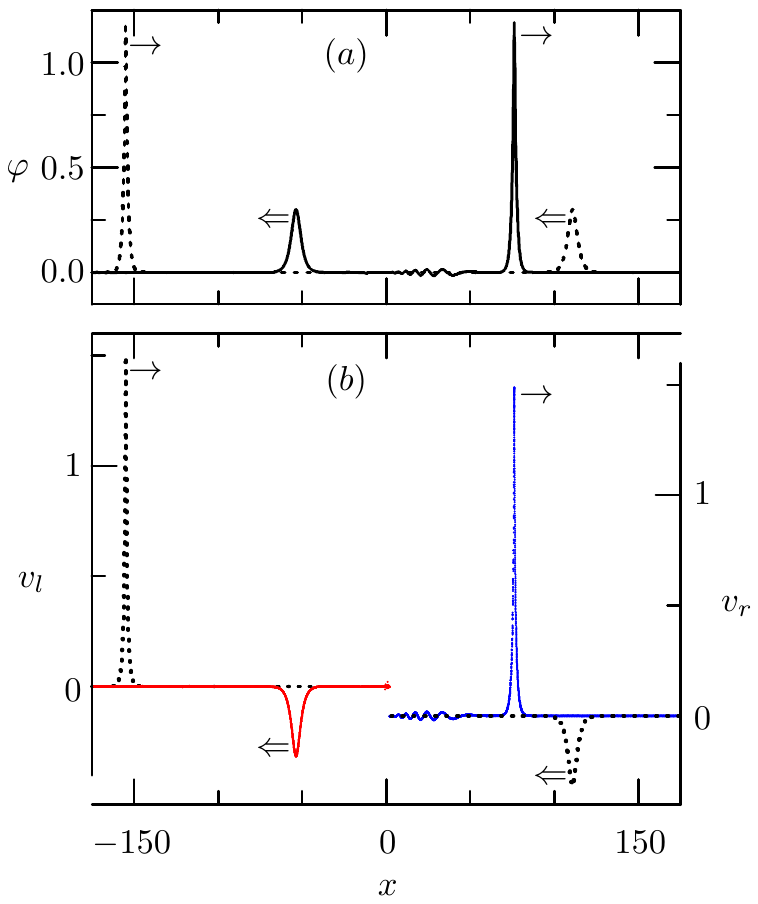}
\caption{The case of   $\varphi_ {m1} ^ 0 = 1.2,  \varphi_ {m2} ^ 0 = 0.3$.
 $(a)$ The distribution of the potential $ \varphi (x) $ at $ t = 0 $ (dotted curves) and $ t = 150 $ (solid curves).  $(b)$
 The phase planes of $l-$ions $(x, v_l)$ (the left curves and the left axis) and the phase planes of $r-$ions $(x, v_r)$  (the right curves and the right axis) at $t = 0$ (dashed curves) and at $t = 150$ (dots, red on-line for $l-$ions and blue on-line for $r-$ions).
}
\label{Fig2}
\end{figure}

In the first case, the amplitudes of the potential of both solitary waves are  $\varphi_ {m1} ^ 0 = \varphi_ {m2} ^ 0 = 0.8 $ (Fig.1).
We can see that  the solitary waves  preserve their shapes, the potential amplitudes, and the ion velocity amplitudes after the collision.
The solitary waves of this amplitude are solitons.
Note that the amplitude range of the potential of solitary waves in a plasma with cold ions and Boltzmann electrons is limited to 1.256 (an approximate estimate of the maximum critical value of the potential amplitude $\varphi_{max}=1.3$ was first given by Sagdeev \cite{Sagdeev}).
The second case corresponds to the collision of solitary  wave 1, whose initial amplitude is close to critical one,  $\varphi_{m1}^0=1.2$, with solitary  wave 2 of a relatively small initial amplitude,   $\varphi_{m2}^0=0.3$, that is, with a soliton.
As can be seen from Fig. 2, in this case also every solitary wave  preserves its identity after the collision.

However, in the other two cases, the situation after the collision of solitary waves is different from the situation presented above.
In the third case, the collision of solitary waves with the initial amplitudes $\varphi_ {m1} ^ 0 = 1.05$  and $\varphi_ {m2} ^ 0 = 0.8$ is considered.
Note that here the initial amplitudes of solitary waves differ noticeably from the critical one.
It can be seen from Fig. 3 that the amplitude and shape of solitary wave 1, the solitary wave with a larger initial amplitude, change sharply after the collision.
The amplitude of solitary wave 1 after the collision falls even lower than the amplitude of solitary wave 2 remaining at the previous level.
The decrease in the amplitude is due to the fact that after the collision of the solitary waves a group of accelerated  ions $b1$ appeared (Fig. 3b).
The ions are accelerated at the expense of the energy of solitary wave 1.
The loss of energy by solitary wave 1 leads to a decrease in its amplitude.
Here, the initial amplitude of solitary wave 1 is less than the initial amplitude of solitary wave 1 in the previous case.
Despite this, solitary wave 1 does not preserve its identity after the collision.
Obviously, this is due to the fact that in this case the initial amplitude of solitary wave 2 is greater than in the previous case.
It is particularly remarkable that solitary wave 2 preserves its shape, the potential amplitude (Fig. 3a) and the ion velocity amplitude (Fig. 3b) after the collision. 

\begin{figure}\centering
\includegraphics[viewport= 209 528 424 784, width=215pt]{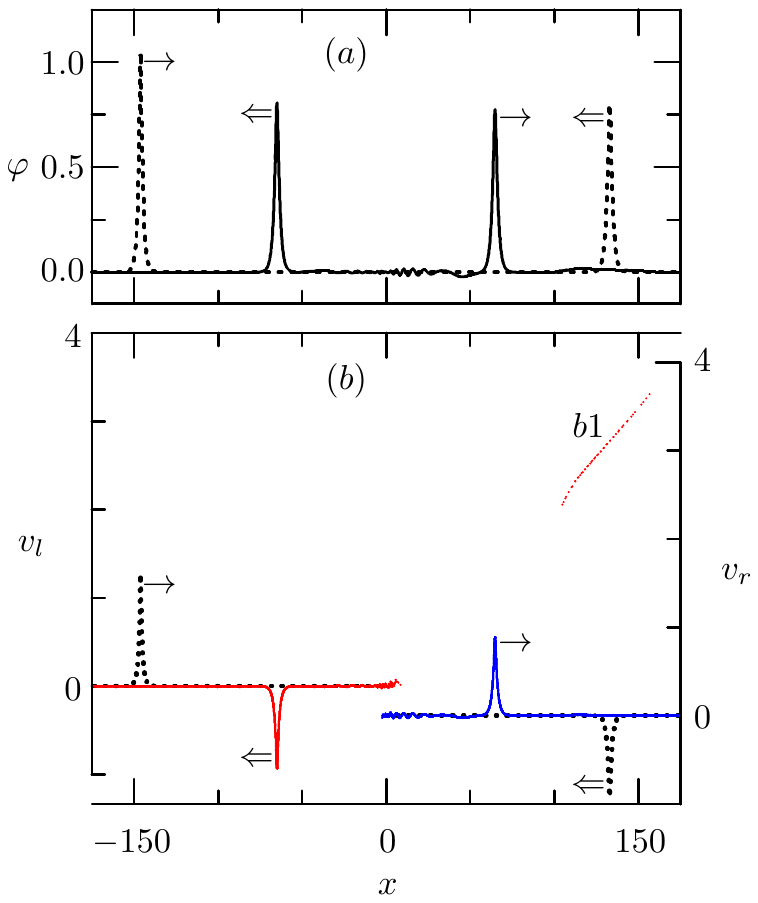}
\caption{The case of  $\varphi_ {m1} ^ 0 =1.05,  \varphi_ {m2} ^ 0 = 0.8$.
 $(a)$ The distribution of the potential $ \varphi (x) $ at $ t = 0 $ (dotted curves) and $ t = 150 $ (solid curves). $(b)$
 The phase planes of $l-$ions $(x, v_l)$ (the left curves  and  curve $b1$, the left axis) and the phase planes of $r-$ions $(x, v_r)$  (the right curves and   the right axis) at $t = 0$ (dashed curves) and at $t = 150$ (dots, red on-line for $l-$ions and blue on-line for $r-$ions).}
\label{Fig3}
\end{figure}

In the last fourth case presented in Fig. 4, both colliding solitary waves have the same initial potential amplitude $\varphi_ {m1} ^ 0 = \varphi_ {m2} ^ 0 = 1.05$.
 As we see, these amplitudes are not preserved after the collision (Fig. 4a).
The initial amplitudes of the ion velocity are also not preserved (Fig. 4b).
In the collision, solitary wave 1 gives a part of its energy to the bunch of accelerated $l-$ions $b1$, and a part of the energy of solitary wave 2 is expended to form a bunch of accelerated $r-$ions $b2$.
This result supports the conclusion  that both solitary waves do not preserve their amplitudes and shapes after a mutual head-on collision if their initial amplitudes are sufficiently large \cite {PhysPlaRep}.
Moreover, in Ref. 29, 
 it was shown that the amplitudes of the potential and the velocity of ions of a similar solitary wave after a collision fall with an increase in the initial amplitude of a counter solitary wave.
We see that such solitary waves are not solitons.

\begin{figure}\centering
\includegraphics[viewport= 209 528 424 784, width=215pt]{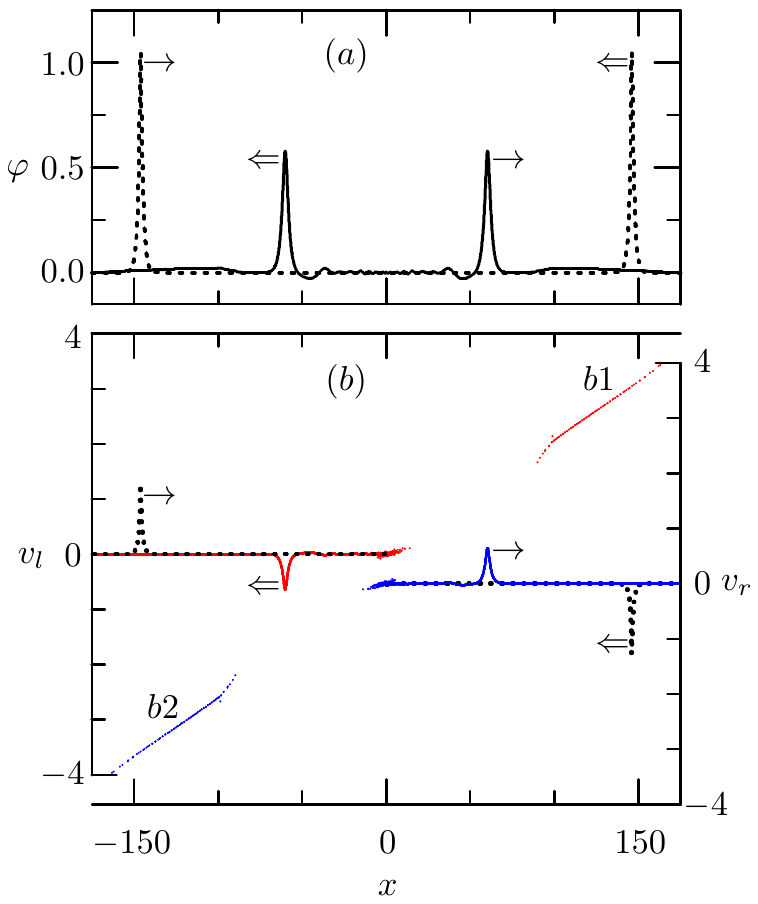}
\caption{The case of  $\varphi_ {m1} ^ 0 =1.05,  \varphi_ {m2} ^ 0 = 1.05$.
 $ (a) $ The distribution of the potential $ \varphi (x) $ at $ t = 0 $ (dotted curves) and $ t = 150 $ (solid curves). 
 $(b)$ The phase planes of $l-$ions $(x, v_l)$ (the left curves and  curve $b1$, the left axis) and the phase planes of $r-$ions $(x, v_r)$  (the right curves and  curve $b2$, the right axis) at $t = 0$ (dashed curves) and at $t = 150$ (dots, red on-line for $l-$ions and blue on-line for $r-$ions). }
\label{Fig4}
\end{figure}

Thus, we briefly described four possible scenarios of a head-on collision of solitary waves.
The possibility of developing one or another scenario depends on the initial amplitudes of both solitary waves.
Two values of these amplitudes can be represented by a point on the plane  of all possible initial amplitudes of two colliding solitary waves  $(\varphi_ {m1} ^ 0, \varphi_ {m2} ^ 0)$.
Each point on this plane corresponds to the collision of solitary waves with initial amplitudes determined by two coordinates of this point.
It is obvious that in the diagram  $(\varphi_ {m1} ^ 0, \varphi_ {m2} ^ 0)$ several characteristic regions corresponding to different scenarios of collision of solitary waves can be distinguished.
We carried out a large number of numerical experiments on the collision of solitary waves to determine all possible characteristic regions and their boundaries in the diagram  $(\varphi_ {m1} ^ 0, \varphi_ {m2} ^ 0)$.
The results of these numerical experiments are shown in Fig. 5.
In the diagram $(\varphi_ {m1} ^ 0, \varphi_ {m2} ^ 0)$, the range of initial amplitudes of each of the colliding waves includes all possible values up to the critical value [0, 1.256].
Obviously, this diagram is symmetric about the straight line from the point (0,0) to the point (1.256, 1.256).

To determine any point on the  boundary between two characteristic regions, it is necessary to perform several numerical experiments.
In all these numerical experiments, the initial amplitude of one of the solitary waves, for example, $\varphi_ {m1} ^ 0$, is the same, but the initial amplitudes of the second solitary wave $\varphi_ {m2} ^ 0$ differ from each other by 0.025 when passing from one numerical experiment to another.
Performing numerical experiments, one can find two cases of collisions of solitary waves, which occur according to different scenarios, and the values of the initial amplitude $\varphi_ {m2} ^ 0$ in both cases differ by the value of 0.025.
The average value of these two initial amplitudes is taken as the value of the initial amplitude of the second solitary wave  $\varphi_ {m2} ^ 0$, which, together with the specified value $\varphi_ {m1} ^ 0$,  determines the desired point on the  boundary between the characteristic regions in the diagram  $(\varphi_ {m1} ^ 0, \varphi_ {m2} ^ 0)$.
Obviously, the error in determining the boundaries between regions does not exceed 0.025 in absolute value.
Having determined all the characteristic regions and their boundaries in the diagram $(\varphi_ {m1} ^ 0, \varphi_ {m2} ^ 0)$, we thereby, in essence, classify the head-on collisions of solitary waves.
We can also find the range of amplitudes, in which solitons exist in such a plasma.

\begin{figure}\centering
\includegraphics[viewport= 190 571 412 784, width=222pt]{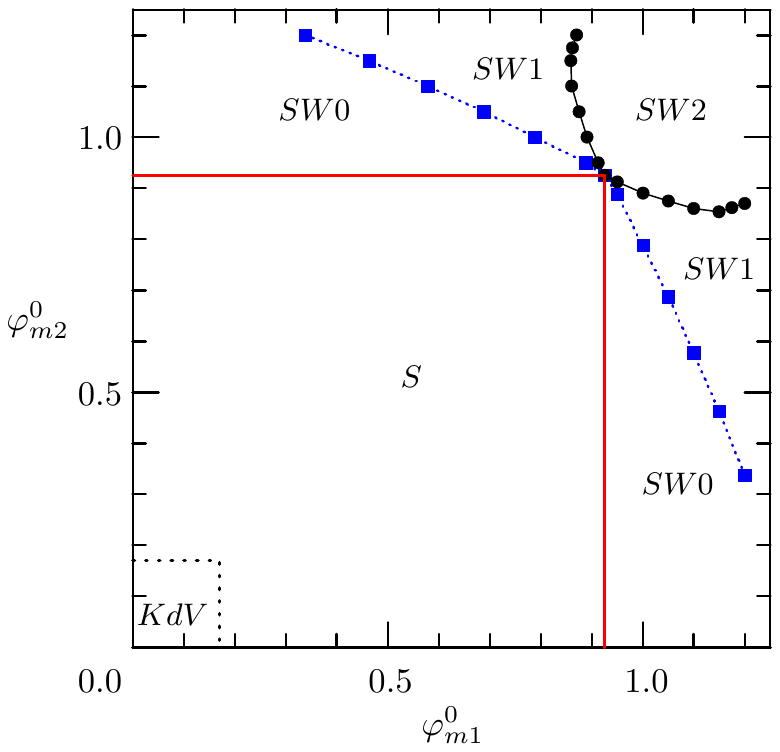}
\caption{The regions of solitons $KdV$ and $S$ and the regions of solitary waves $SW0$, $SW1$ and $SW2$ in the diagram $(\varphi_{m1}^0, \varphi_{m2}^0)$.
}
\label{Fig5}
\end{figure}

As we have already mentioned, a solitary wave, whose amplitude is less than 0.17, can be called the KdV soliton.
Therefore, here the region of the diagram $(\varphi_ {m1}^0, \varphi_ {m2}^0)$, where each point corresponds to a collision of solitary waves with initial potential amplitudes less than 0.17, is referred to as the region of KdB solitons.
Fig. 5 shows that this region, designated as KdV, occupies only a  small part of the diagram $(\varphi_ {m1}^0, \varphi_ {m2}^0)$.
A much larger part of the diagram is occupied by the region denoted by the letter S.
This region corresponds to collisions of solitary waves with initial potential amplitudes in the range [0.17, 0.925].
Here, solitary waves preserve their amplitudes and shapes after a collision, that is, they are solitons.
 These solitons can not be described by the solution of the KdV equation.
The collision of such solitons was illustrated above in Fig. 1.
From our consideration it follows that a soliton is a solitary wave with an amplitude of potential less than 0.925.

Solitary waves with potential amplitudes greater than 0.925 are not solitons.
From numerical experiments, it follows that when such solitary waves collide with other solitary waves or solitons, three scenarios are possible.
In the diagram $(\varphi_ {m1}^0, \varphi_ {m2}^0)$, we can see the regions $SW0$, $SW1$ and $SW2$ that correspond to these scenarios.
If the initial amplitudes of the potential of the colliding solitary waves are such that the corresponding point on the diagram  $(\varphi_ {m1}^0, \varphi_ {m2}^0)$ lies in one of the two regions labeled $SW0$, then both solitary waves preserve  their amplitudes and shapes after the collision.
Here, a solitary wave of large initial amplitude, which is not a soliton, behaves like a soliton in a collision with a soliton of a sufficiently small amplitude.
This situation is illustrated by the case presented in Fig. 2.

If in a collision between a solitary wave and a soliton the amplitude of the soliton is not small, and the point corresponding to the collision is in one of the two regions $SW1$ on the diagram $(\varphi_ {m1}^0, \varphi_ {m2}^0)$, then the solitary wave does not preserve its identity after the collision.
However, the amplitude and shape of the soliton remain unchanged.
A collision of this type has been illustrated above in Fig. 3.
At large initial amplitudes of both colliding solitary waves, their identities are not preserved after the collision.
This occurs when the initial amplitudes of the colliding solitary waves are such that the corresponding point on the diagram  $(\varphi_ {m1}^0, \varphi_ {m2}^0)$ is in the region $SW2$.
Above, we saw this type of collision of solitary waves in Fig. 4.

Thus, we have described regions that differ from each other in the diagram $(\varphi_ {m1}^0, \varphi_ {m2}^0)$.
Each region is characterized by the fact that all possible collisions of two solitary waves, the initial amplitudes of which are represented by a point in this region, occur according to the same scenario.   
 Different regions correspond to different scenarios.
An exception is the difference between the soliton regions $KdV$ and $S$.
This difference consists only in the possibility or impossibility of an analytical description of solitons.
In other regions of the diagram $(\varphi_ {m1}^0, \varphi_ {m2}^0)$, at least one of the colliding solitary waves is not a soliton. 
Thus, using the diagram  $(\varphi_ {m1}^0, \varphi_ {m2}^0)$, we have classified head-on collisions of solitary waves in a plasma with cold ions and Boltzmann electrons.
With the help of this diagram, we can predict the scenario of the collision of two solitary waves, knowing their initial amplitudes.

\section{CONCLUSIONS}

In the present work, using numerical simulation by the particles-in-cell method, we investigate head-on collisions of solitary waves in a plasma with cold ions and Boltzmann electrons.
It is shown that the collision of solitary waves can occur under different scenarios, depending on the initial amplitudes of the colliding solitary waves. 
Examples of four different scenarios, in which each solitary wave either preserve its identity after a collision or do not preserve it, are given.

Possible scenarios of collisions of solitary waves are determined by their initial amplitudes.
Two initial amplitudes of colliding solitary waves can be taken as the coordinates of a point on the plane  of all possible initial amplitudes of two colliding solitary waves  $(\varphi_ {m1} ^ 0, \varphi_ {m2} ^ 0)$.
Each point in the diagram  $(\varphi_ {m1} ^ 0, \varphi_ {m2} ^ 0)$ corresponds to the collision of solitary waves with initial amplitudes determined by two coordinates of this point.
In the diagram  $(\varphi_ {m1} ^ 0, \varphi_ {m2} ^ 0)$, one can find several characteristic regions corresponding to different scenarios of collision of solitary waves.

There are two soliton regions, i.e. the regions of the diagram $(\varphi_ {m1} ^ 0, \varphi_ {m2} ^ 0)$ corresponding to such collisions, in which both solitary waves preserve their amplitudes and shapes after the collision.
These regions differ from each other in the range of initial amplitudes of solitary waves.
In the region of small amplitudes $\varphi_ {m} ^ 0 \le 0.17$, solitons can be described by a solution of the KdV equation.
But in the second soliton region $[0.17, 0.925]$, the initial amplitudes of the potential are sufficiently large and, therefore, using the solution of the KdV equation to describe solitons can lead to large errors.
Because of this, it makes sense to distinguish between two soliton regions.

A solitary wave with an amplitude greater than 0.925, strictly speaking, is not a soliton.
But in the diagram  $(\varphi_ {m1} ^ 0, \varphi_ {m2} ^ 0)$ there is a region $SW0$, corresponding to the collisions of such solitary waves with solitons, after which the solitary waves remain unchanged.
In these collisions solitary waves that are not solitons behave like solitons in collisions.
In the diagram  $(\varphi_ {m1} ^ 0, \varphi_ {m2} ^ 0)$, there is another region  $SW1$, also corresponding to  collisions of solitary waves with solitons.
But in these collisions,  solitary waves do not preserve their identities after a collision, while solitons remain unchanged.
Finally, the region of very large amplitudes of  both colliding solitary waves  $SW2$ corresponds to collisions, at which both solitary waves lose their identities.
In essence, the diagram  $(\varphi_ {m1} ^ 0, \varphi_ {m2} ^ 0)$  classifies collisions of solitary waves.
Using this diagram, it is easy to predict the scenario, in which the solitary waves will collide.

\end{document}